# Anomalous magnetic response in the Au–Al–Gd 1/1 quasicrystal approximant

Takafumi D. Yamamoto[1], Tasuku Watanabe[1], Farid Labib[2], Asuka Ishikawa[2], Shintaro Suzuki[3], and Ryuji Tamura[1]

[1] *Department of Materials Science and Technology, Tokyo University of Science, Tokyo 125-8585, Japan*
[2] *Research Institute of Science and Technology, Tokyo University of Science, Tokyo 125-8585, Japan*
[3] *Department of Physical Science, Aoyama Gakuin University, Sagamihara 252-5258, Japan*

Corresponding Author: td_yamamoto@rs.tus.ac.jp

**Abstract:**

The magnetic response of the Tsai-type 1/1 Au–Al–Gd approximant crystals (ACs) was quantitatively investigated in terms of the magnetic entropy change ($\Delta S_M$) for different magnetic ground states. A comprehensive $\Delta S_M$ map over a wide electron concentration range has been established, demonstrating the detailed variation of $\Delta S_M$ across the entire magnetic phase diagram in the Tsai-type 1/1 ACs. Near the boundaries of the ferromagnetic (FM) phase, a clear deviation from the mean-field theory (MFT) was observed in both the Curie temperature ($T_C$) and magnetic field ($H$) dependences of the maximum magnetic entropy change ($\Delta S_M^{\max}$). Contrary to general expectations, a high $\Delta S_M^{\max}$ (7.2 J/K mol-Gd under a 5 T field variation), even comparable to those of candidate materials for low-temperature magnetic refrigeration, was obtained within the antiferromagnetic (AFM) region near the FM / AFM phase boundary. The unexpected enhancement of $\Delta S_M^{\max}$ toward the AFM region under high magnetic fields indicates an anomalous magnetic response in the present Tsai-type 1/1 AC, which is presumably associated with the breakdown of the MFT. The present finding suggests that tuning the magnetic ground state across the phase boundary is an effective strategy to enhance $\Delta S_M$, even in general rare-earth intermetallic compounds.

# I. INTRODUCTION

Long-range magnetic order in quasicrystals (QCs) and their approximant crystals (ACs) has been one of the fascinating topics in condensed matter physics, as novel physical properties are expected to emerge reflecting their unique long-range as well as short-range icosahedral atomic arrangements. Since the first discovery of an antiferromagnetic order in $Cd_6Tb$ 1/1 AC [1], various magnetic orders were discovered in numerous rare-earth (RE)–bearing Tsai-type ACs [2-12], and recently, ferromagnetic (FM) and antiferromagnetic (AFM) orders were finally discovered in Tsai-type icosahedral QCs [13-15]. Understanding of the magnetism in these systems has also made remarkable progress, especially in the Au–SM (SM: Semimetal)–RE 1/1 ACs with a body-centered-cubic lattice (see Fig. S1 in the Supplemental Material [16]) and a wide single-phase stability region in the phase diagram. One of the significant findings is the continuous variation of the net magnetic interaction with the electron-to-atom ratio ($e/a$) [8, 16-19], which allows for continuous change in magnetic ground state and magnetic transition temperature ($T_{mag}$) within an isostructural single-phase system. Furthermore, recent neutron diffraction experiments [20-24] have unveiled unique non-coplanar spin configurations on magnetically coupled icosahedral RE-clusters.

Lately, Kikugawa et al. [25] discovered considerable isothermal magnetic entropy changes ($\Delta S_M$) in FM Au–Al–RE (RE = Gd, Tb, Dy) 1/1 ACs. This quantity, closely related to the magnetic response of magnetic moments, is expected to reflect peculiar RE-spin arrangements of Tsai-type 1/1 ACs, though the origin of the large $\Delta S_M$ remains unclear. In order to get a deeper understanding of the relation between $\Delta S_M$ and magnetism of Tsai-type 1/1 ACs, it would be of particular importance to reveal how the $\Delta S_M$ of Tsai-type 1/1 ACs depends on the magnetic ground state and $T_{mag}$ through a wide variation of the $e/a$ ratio, which has not been performed to date. For this purpose, Au–Al–Gd 1/1 AC, one of the most studied alloys [7, 8, 26-29], is an ideal system to be investigated because (i) it has the largest single-phase region in the $e/a$ parameter space, which allows the magnetic ground state to change from spin-glass (SG) to FM and further to AFM states; and (ii) the FM region is particularly wide allowing to control the Curie temperature over a wide $e/a$ range.

In this study, we present the first example of a fully comprehensive $\Delta S_M$ map with widely varying $e/a$ ratio over $e/a$ = 1.54 – 1.98 in Au–Al–Gd 1/1 ACs. We find that the maximum $\Delta S_M$ undergoes a sharp rise below $e/a$ = 1.87, corresponding to a border between FM and SG regions in the $e/a$ parameter space. Within the former region where $e/a$ = 1.60 – 1.86, the $\Delta S_M$ value remains nearly unchanged despite the variation in Curie temperature from 10 to 30 K, in contrast to the often-observed trend in a wide range of ferromagnets. Moreover, below $e/a$ = 1.59, the maximum $\Delta S_M$ experiences an unexpected enhancement toward the AFM region near the FM / AFM border under high magnetic fields, reaching ~7.2 J/K mol-Gd for a 5T field variation, a value comparable to those of candidate materials for low-temperature magnetic refrigeration. The variation of $\Delta S_M$ versus $e/a$ ratio in the FM regime is discussed within the framework of the mean-field theory (MFT), demonstrating a clear breakdown of the MFT near the phase boundaries in the Tsai-type 1/1 AC.

# II. EXPERIMENTAL

Polycrystalline 1/1 ACs with nominal compositions of $Au_xAl_{86-x}Gd_{14}$ ($x$ = 51 – 73) were synthesized by arc-melting technique followed by heat treatment at 1073 K for 50 hours under an argon atmosphere. Powder X-ray diffraction patterns at 300 K were acquired using a Rigaku MiniFlex600 diffractometer with Cu-$K\alpha$ radiation. The results confirm that all the samples are single phase of Tsai-type Au–Al–Gd 1/1 AC (see Fig. 1). Moreover, as shown in Fig. S2 [16], the lattice constant increases systematically with increasing the nominal Au content $x$, indicating a mutual substitution of smaller Al atoms by larger Au atoms. The chemical composition of the samples was examined by a JEOL JSM-IT100 scanning electron microscope equipped with energy-dispersive X-ray spectrometer. A systematic variation in chemical composition of the samples with the Au substitution for Al is indicated in Table S1, which is also evident from Fig. S3 where the analyzed Au content increases linearly with increasing $x$ [16]. Magnetization ($M$) were measured in the temperature ($T$) range of 2 – 100 K under external magnetic fields ($\mu_0H$) ranging from 0 to 5 T using a superconducting quantum interference device magnetometer (Quantum Design, MPMS3).

## III. RESULTS AND DISCUSSION

Figures 2(a)–2(c) provide iso-field magnetization ($M$-$T$) curves of the $Au_xAl_{86-x}Gd_{14}$ 1/1 ACs with nominal Au content $x = 51 - 73$ at 0.01 T in zero-field cooling (ZFC) and field cooling (FC) processes. Figure 2(d) depicts isothermal magnetization ($M$-$H$) curves at 2 K for selected samples with $x = 51$, 62, and 73. Clearly, three different magnetic behaviors can be distinguished over a wide $x$ range. For $x < 56$ [Fig. 2(a)], small cusps appear in ZFC curves around $T = 3 - 4$ K, below which FC curves start to bifurcate, as evident in the inset of Fig. 2(a). Within this region, $M$ at 2 K reaches only 2 $\mu_B$/Gd at 5 T [Fig. 2(d)], much below the full moment of a free $Gd^{3+}$ ion (7 $\mu_B$/Gd). These results are consistent with typical SG-like behaviors [30]. At intermediate $x$-range of 57 – 70, the $M$-$T$ curve increases steeply below a certain temperature [Fig. 2(b)] and $M$ at 2 K rapidly saturates toward ~7 $\mu_B$/Gd [Fig. 2(d)], a behavior expected for FM compounds. Above $x = 72.5$, the occurrence of an AFM transition is evidenced by the appearance of clear sharp cusps in both ZFC and FC curves around 9.5 K [see the inset of Fig. 2(c)] as well as the metamagnetic anomaly around 0.5 T at 2 K [Fig. 2(d) and its inset].

The above results allow us to determine the magnetic phase diagram of the $Au_xAl_{86-x}Gd_{14}$ 1/1 ACs with respect to the $e/a$ ratio under the assumption of Au being monovalent and Al, Gd being trivalent. The obtained $T_{mag}$ versus $e/a$ phase diagram is presented in Figure 3, where $T_{mag}$ represents the Néel temperature $T_N$, the Curie temperature $T_C$, and the spin freezing temperature $T_f$ for the AFM, FM, and SG states, respectively. A similar magnetic phase diagram has also been reported elsewhere [8]. The figure clearly shows the systematic changes of magnetic ground state and $T_{mag}$ with the $e/a$ ratio in the Au–Al–Gd 1/1 AC. Notably, the FM state is realized over a wide $e/a$ range accompanying with a significant change of $T_C$, which provides an ideal opportunity to examine $T_C$-dependence of $\Delta S_M$ within an isostructural single-phase system without disturbing the network of magnetic RE ions.

In the following, we investigate the $\Delta S_M$ of the present 1/1 ACs from a series of $M$-$T$ curves in FC process recorded at various magnetic fields up to 5 T; the examples of which are shown in Fig. S4 for representative samples of each magnetic ground state [16]. For the $\Delta S_M$ calculation, the following thermodynamic Maxwell's relation is utilized:

$$\Delta S_M(T,\mu_0 H) = \mu_0 \int_0^H \left(\frac{\partial M(T,\mu_0 H)}{\partial T}\right)_H dH . \tag{1}$$

A negative value of this quantity indicates an entropy release upon application of a magnetic field. The temperature dependence of $\Delta S_M$ for $\mu_0 H$ = 5 T in all the samples are provided in Fig. 4, showing that the maximum magnetic entropy change $\Delta S_M^{max}$ is observed around $T_{mag}$ in each sample, except for the sample with $e/a$ = 1.88, located at the SG / FM phase boundary (see Fig. 3). Note here that the $\Delta S_M^{max}$, or $\Delta S_M (T_{mag}, \mu_0 H)$, is a function of two variables, i.e., $T_C$ and $\mu_0 H$, in the FM region ($e/a$ = 1.60 – 1.86), and henceforth the dependence of $\Delta S_M^{max}$ on both variables will be separately examined below.

First, we discuss the $T_C$ dependence of the $\Delta S_M^{max}$ in the FM Au–Al–Gd 1/1 ACs as shown in Fig. 5(a). Surprisingly, $\Delta S_M^{max}$ is found to be almost independent on $T_C$ over the wide FM region despite $T_C$ largely varies between 10 – 30 K, exhibiting an unconventional behavior that goes against the trend often observed in a wide range of ferromagnets, where the $\Delta S_M^{max}$ for a given $\mu_0 \Delta H$ decreases with increasing $T_C$ [31-33] (refer to, e.g., Fig. 5 of Ref [33]). From the Weiss mean field equation of state and the Taylor expansion of the Brillouin function ($B_J$), the well-established trend of $\Delta S_M^{max}$ at $T_C$ is given as [34],

$$-\Delta S_M^{max} (T_C, \mu_0 H) = \frac{1}{2C}\left(\frac{C^2}{K}\right)^{2/3}\left(\frac{\mu_0 H}{T_C}\right)^{2/3} + \cdots , \qquad (2)$$

indicating $T_C^{-2/3}$ dependence of $\Delta S_M^{max}$, where $C$ is the Curie constant and $K$ is also a constant, both depending only on the total angular momentum $J$. The $T_C^{-2/3}$ dependence predicted from the MFT is also plotted in Fig. 5(a). Clearly, the $\Delta S_M^{max}$ in the present FM 1/1 ACs does not follow this $T_C^{-2/3}$ trend: It remains nearly unchanged or even slightly increases with raising $T_C$, breaking the above law. This discrepancy cannot be attributed to the approximation used in deriving Eq. (2), but is simply due to the breakdown of the validity of the $B_J$ function, from which Eq. (2) is derived. According to the $B_J$ form, the $M$-$H$ curve at $T_C$, *i.e.*, $M (T_C, \mu_0 H)$, is solely dependent on $\mu_0 H/T_C$ for the same RE system (for the same $J$) [34], and therefore $M (T_C, \mu_0 H)$, or consequently $\Delta S_M (T_C, \mu_0 H)$, should inevitably decrease by increasing $T_C$. Thus, it is obvious that the observed $T_C$ dependence of the $\Delta S_M^{max}$ is not compatible with the MFT.

Second, we discuss the magnetic response of the Au–Al–Gd 1/1 ACs in terms of the field dependence of $\Delta S_M^{max}$. To this end, we thoroughly examined it for the FM samples with different $T_C$'s, as summarized in Fig. S5 in the Supplemental Material [16]. While all the $\Delta S_M^{max}$, i.e., $\Delta S_M (T_C, \mu_0 H)$ follow a power-law dependence on the magnetic field ($H^n$), the exponent $n$ was found to range from 0.65 – 0.88. Notably, as represented by the broken curve in Fig. 5(b), a systematic deviation of $n$ from the mean-field value ($n = 2/3$) was recognized when the system moves from the center of the FM region (i.e., $e/a$ ~ 1.70) towards both borders ($e/a$ ~ 1.60 and 1.86), i.e., when $T_C$ decreases.

Both the $T_C$ and $H$ dependences of $\Delta S_M^{max}$ demonstrate a clear breakdown of the MFT in the Au–Al–Gd 1/1 ACs, which becomes pronounced near the boundaries of the FM phase. This is in contrast to previous studies [14, 19, 35, 36], which focused solely on a limited $e/a$ range within the FM regime and concluded the mean-field behavior of Tsai-type compounds. It is generally understood that the MFT is not valid in the presence of significant spin fluctuations, as its major assumption is the neglection of spin fluctuations. Accordingly, enhanced spin fluctuations near the magnetic phase boundaries may be one of the possible reasons for the MFT breakdown observed in the present system. The precise origin of which requires further experimental verifications, such as direct observation by nuclear magnetic resonance experiments.

Figure 6(a) illustrates the comprehensive $\Delta S_M^{\text{max}}$ map as a function of $e/a$ ratio and $\mu_0 H$, visually demonstrating an increase in $\Delta S_M^{\text{max}}$ near the FM / AFM phase boundary ($e/a \sim 1.59$). Indeed, the magnitude of $\Delta S_M^{\text{max}}$ for $\mu_0 H$ = 5 T experiences a clear enhancement from ~5.7 J/K mol-Gd below $e/a \sim 1.59$, reaching ~7.2 J/K mol-Gd within the AFM region [see Fig. 6(b)]. This $\Delta S_M^{\text{max}}$ value is about 1.4 – 1.9 times larger than those of similar Tsai-type 1/1 ACs [25, 35, 36], marking the highest $\Delta S_M$ value ever reported in Tsai-type compounds to date. Furthermore, this value is even comparable to or larger than those of some RE-based compounds having a similar $T_{\text{mag}}$ of 10 K [33, 37-40]. Note that not many magnetic materials, even general RE-based compounds, have the $\Delta S_M$ values exceeding 6 J/K mol-RE for $\mu_0 H$ = 5 T in the temperature range of 2 – 40 K [41]. In this respect, the value of ~7.2 J/K mol-Gd represents a significant improvement in the magnetocaloric properties of Tsai-type compounds.

Notably, the largest $\Delta S_M$ occurs within the AFM region, not in the FM region, at high magnetic fields above 2 T. It is widely believed that FM materials are advantageous for large $\Delta S_M^{\text{max}}$ compared to AFM materials [42-44], owing to a steeper rise in magnetization near $T_C$. One may notice that when a high magnetic field is applied at low temperatures, the considerable Zeeman energies may force the magnetic moments of the AFM material to align along the field direction, resulting in a forced FM state and consequently a large $\Delta S_M^{\text{max}}$. However, even taking such cases into account, when comparing materials containing the same RE element, FM materials are expected to exhibit larger $\Delta S_M^{\text{max}}$ at any magnetic field due to greater magnetization [42]. Hence, within the same RE system, the $\Delta S_M^{\text{max}}$ should be larger in FM materials than in AFM materials. This is true for the present system at low magnetic fields, as seen in Fig. 6(b) where the $\Delta S_M^{\text{max}}$ for $\mu_0 H$ = 1 T increases as the system moves into the FM region from the AFM region. Conversely, the observed enhancement of $\Delta S_M^{\text{max}}$ toward the AFM region near the FM / AFM border under high magnetic fields is right opposite to the general expectations and therefore represents an anomalous magnetic response in the present Au–Al–Gd 1/1 AC, presumably associated with the breakdown of the MFT near the FM phase boundaries.

This study demonstrates that the $\Delta S_M$ of the Tsai-type 1/1 ACs is significantly enhanced by tuning the magnetic ground state toward the magnetic phase boundary through controlling $e/a$ ratio. This phenomenon, which has not been recognized before, is expected to occur in other Tsai-type compounds as well. Therefore, the present findings provide a new design principle for new quasicrystal-related compounds with potentially promising magnetocaloric performance. Note that an important role of non-magnetic elements such as Au and Al in Tsai-type compounds is the freedom they provide in tuning the $e/a$ ratio. These sites are, in principle, replaceable by lower-cost elements, enabling the development of more cost-effective alloy systems. Finally, the principle proposed here is not limited to Tsai-type compounds but may also be applicable to general RE intermetallic compounds through similar $e/a$ tuning, provided that their magnetism is governed by the RKKY interaction.

## IV. SUMMARY

In summary, the relationship between magnetic response and electron-to-atom ($e/a$) ratio was investigated in a series of Tsai-type $Au_xAl_{86-x}Gd_{14}$ ($x$ = 51–73) 1/1 approximant crystals (ACs), based on the magnetic entropy change ($\Delta S_M$). The results demonstrated the detailed variation of $\Delta S_M$ across different magnetic phases of the Tsai-type 1/1 ACs. Near the boundaries of the ferromagnetic (FM) phase, a clear breakdown of the mean-field theory (MFT) was observed in both the $T_C$ and $H$ dependences of the maximum magnetic entropy change ($\Delta S_M^{\text{max}}$).

Contrary to general expectations, the $\Delta S_M^{\max}$ was clearly enhanced toward the antiferromagnetic (AFM) region near the FM / AFM phase boundary under high magnetic fields, reaching about 7.2 J/K mol-Gd (under a 5 T field variation), a value comparable to those of other RE compounds considered promising for low-temperature magnetic refrigeration. This phenomenon represents an anomalous magnetic response under high magnetic fields in the present Gd-containing Tsai-type 1/1 ACs, which is presumably associated with the breakdown of the MFT near the FM phase boundaries.

## ACKNOWLEDGMENTS

This work was supported by KAKENHI Grants-in-Aid for Scientific Research (Grant No. JP19H05817, JP19H05818, JP24K17016) from Japan Society for the Promotion of Science and by CREST (Grant No. JPMJCR22O3) from Japan Science and Technology Agency.

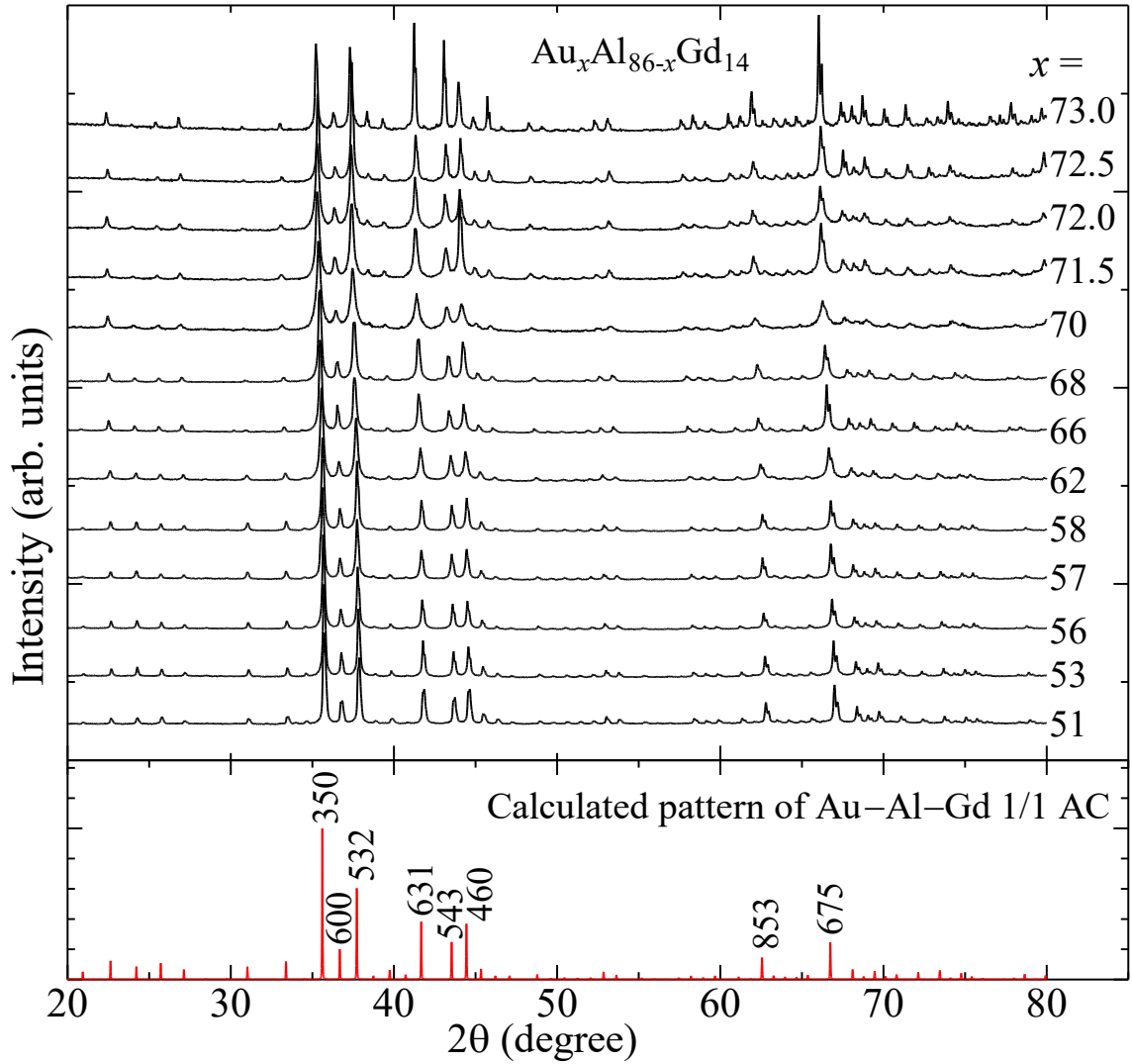

Fig. 1. Powder X-ray diffraction patterns of $Au_xAl_{86-x}Gd_{14}$ 1/1 ACs with nominal composition of $x = 51$–$73$. A calculated pattern of Tsai-type Au–Al–Gd 1/1 AC (analyzed composition: $Au_{61.4}Al_{24.8}Gd_{13.8}$) with a comparable lattice constant [27] is shown on the bottom panel for comparison.

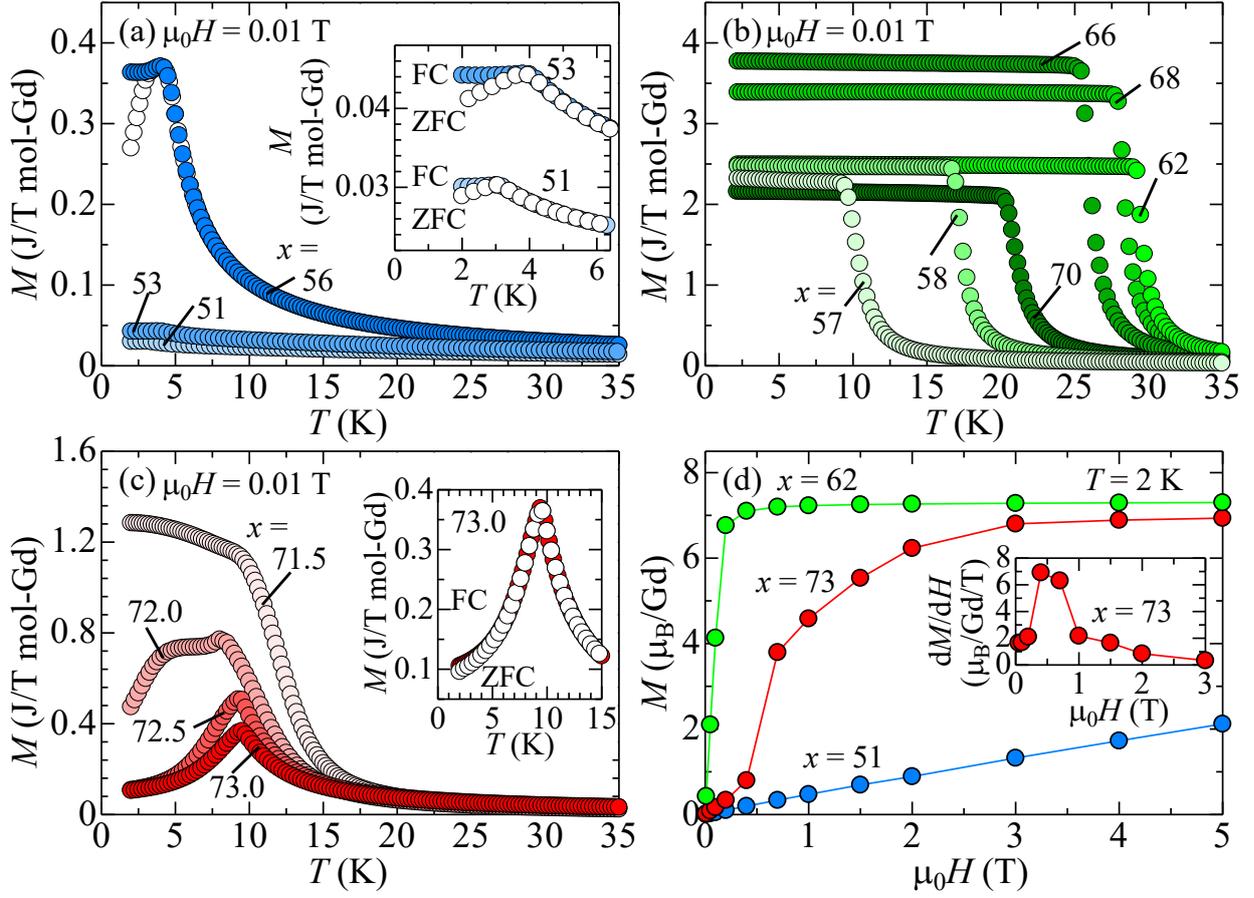

Fig. 2. (a)–(c) M-T curves at 0.01 T for $Au_xAl_{86-x}Gd_{14}$ 1/1 ACs with nominal compositions of (a) $x = 51 – 53$, (b) $x = 57 – 70$, and (c) $x = 71.5 – 73.0$. The insets of (a) and (c) are enlarged views to compare the data for FC and ZFC processes. (d) M-H curves at 2 K for selected samples with $x = 51$, 62, and 73. The inset shows the field derivative of magnetization $dM/dH$ at $x = 73$ as a function of magnetic field below 3 T. The peak anomaly near 0.5 T indicates the occurrence of a metamagnetic transition from an AFM state to a forced FM state.

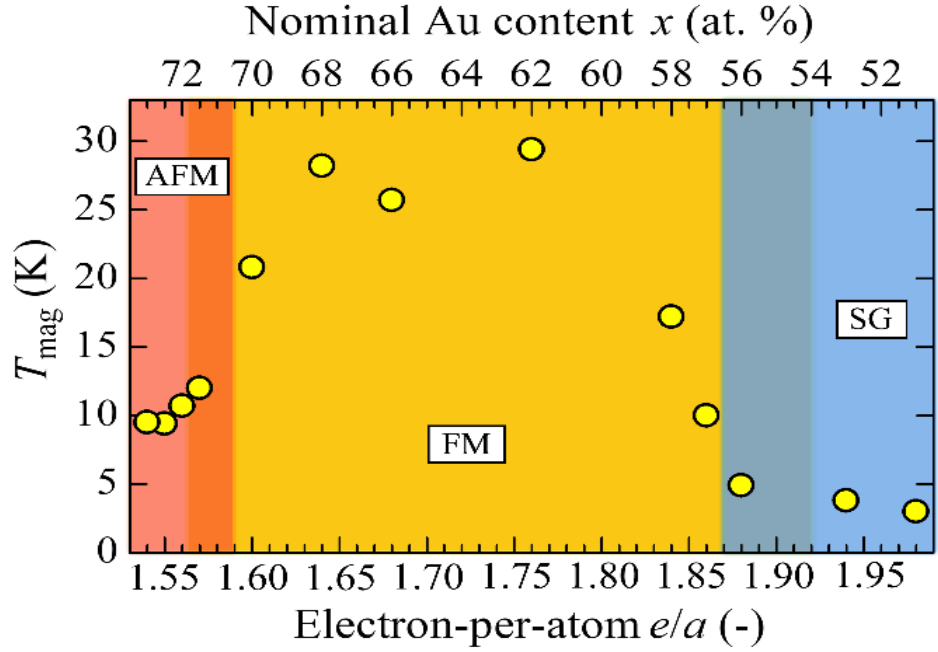

Fig. 3. The variation of $T_{mag}$ versus $e/a$ for Au–Al–Gd 1/1 ACs, along with the magnetic ground states; antiferromagnetic (AFM), ferromagnetic (FM), and spin-glass (SG) states. The upper axis represents the nominal Au content $x$ of the $Au_xAl_{86-x}Gd_{14}$ 1/1 ACs.

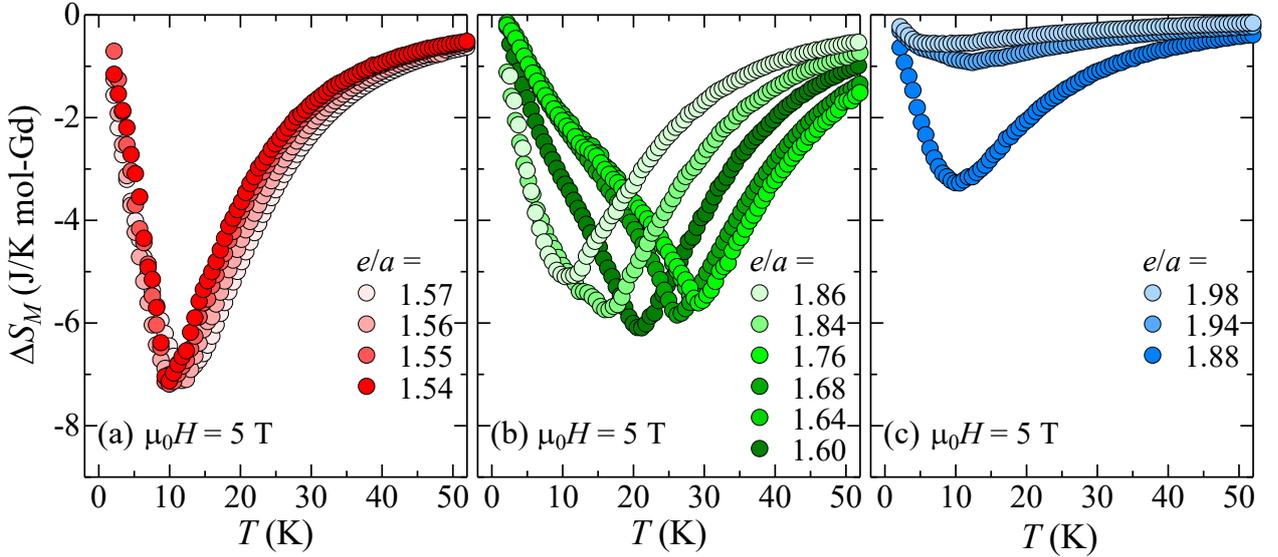

Fig. 4. (a)–(c) Temperature dependence of $\Delta S_M$ for $\mu_0 H = 5$ T in all the samples with the $e/a$ ratio of (a) 1.54 –1.57, (b) 1.60 –1.86, and (c) 1.88–1.98.

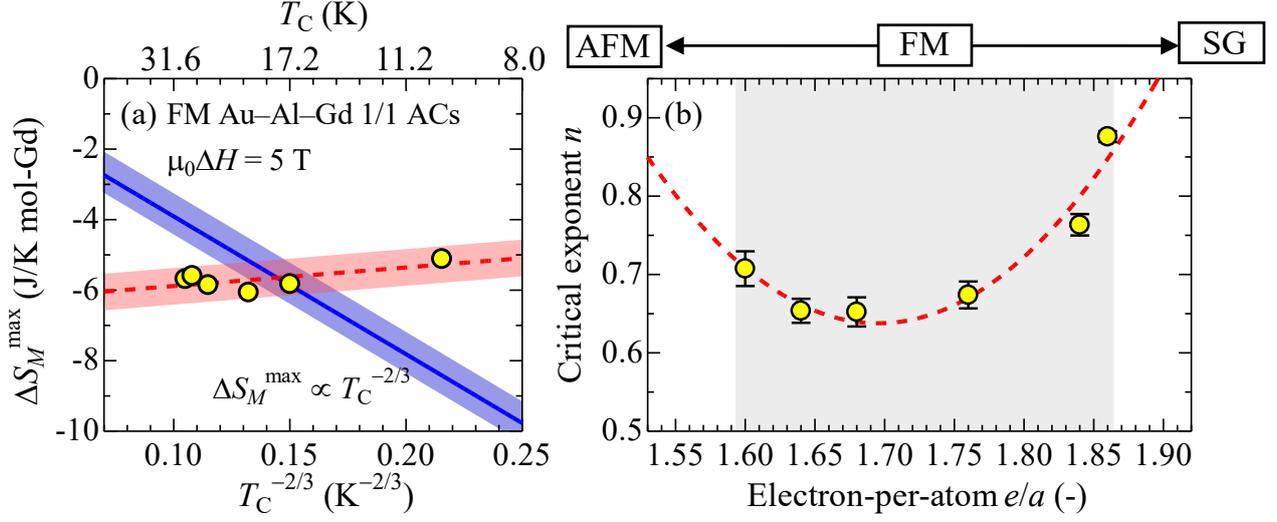

Fig. 5. (a) $\Delta S_M^{max}$ for $\mu_0 \Delta H = 5$ T as a function of $T_C^{-2/3}$ for the FM Au–Al–Gd 1/1 ACs. The solid line represents the $T_C$ dependence of $\Delta S_M^{max}$, predicted from the MFT (see the text). The broken line is the guide to the eye. The upper axis indicates the corresponding $T_C$. (b) The $e/a$ dependence of the critical exponent $n$ derived from the field dependence of $\Delta S_M^{max}$ for the FM 1/1 ACs (see the text). The broken curve is the guide to the eye. The shadow area indicates the FM region.

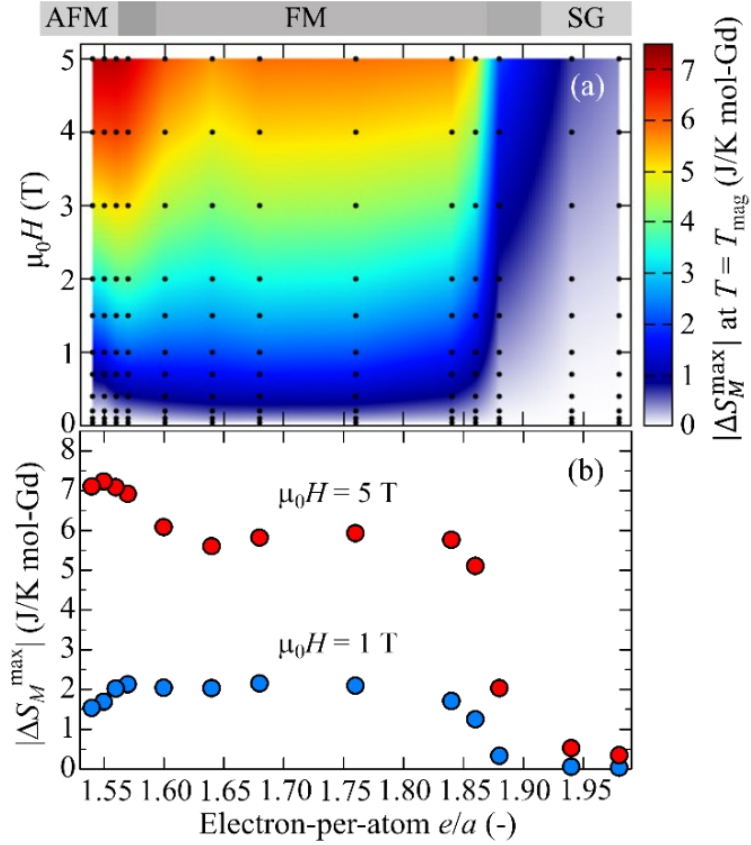

Fig. 6. (a) $|\Delta S_M^{max}|$ map as a function of $\mu_0 H$ and $e/a$ ratio. The black dots represent the data points at respective $\mu_0 H$ and $e/a$ values used to obtain the map. (b) $|\Delta S_M^{max}|$ versus $e/a$ variation under $\mu_0 H = 1$ and 5 T.

# Supplemental Material:
# Anomalous magnetic response in the Au–Al–Gd 1/1 quasicrystal approximant

Takafumi D. Yamamoto[1], Tasuku Watanabe[1], Farid Labib[2], Asuka Ishikawa[2], Shintaro Suzuki[3], Ryuji Tamura[1]

[1]*Department of Materials Science and Technology, Tokyo University of Science, Tokyo 125-8585, Japan*
[2]*Research Institute of Science and Technology, Tokyo University of Science, Tokyo 125-8585, Japan*
[3]*Depeartment of Physical Science, Aoyama Gakuin University, Sagamihara 252-5258, Japan*

**Introduction**

**Fig. S**1 shows the arrangement of rare-earth (RE) atoms in the Tsai-type 1/1 ACs. The structure is described as a body-centered-cubic packing of RE-icosahedral clusters.

**Experimental**

**Fig. S**2 shows the variation of lattice constant *a* with nominal Au content *x* in the synthesized $Au_xAl_{86-x}Gd_{14}$ (*x* = 51–73) samples. The lattice constant increases systematically with increasing *x*, confirming a mutual substitution of smaller Al atoms by larger Au atoms in the samples.

**Fig. S**3 and **Table S**1 shows the analyzed Au content plotted as a function of nominal Au content *x* in the synthesized $Au_xAl_{86-x}Gd_{14}$ samples. The chemical composition analysis was performed by using a scanning electron microscope (SEM) equipped with energy-dispersive X-ray spectrometer (EDX). The observed 2–3 at.% discrepancy between the analyzed and nominal contents, especially for Au and Gd elements, is within the accuracy limit of the SEM-EDX elemental analysis, 5 at.% of the correct value, for a polished bulk target [D. E. Newbury and N. W. M. Ritchie, Scanning **35**, 141 (2013). ].

**Results and discussion**

**Fig. S**4 shows iso-field magnetization (*M-T*) curves measured at various magnetic fields up to 5 T in field cooling (FC) processes for three selected samples with *e*/*a* = 1.98, 1.76, and 1.54. A series of these curves are used to calculate the magnetic entropy change ($\Delta S_M$) through a Maxwell's relation.

**Fig. S**5 shows the magnitude of the maximum magnetic entropy change $|\Delta S_M^{\max}|$ as a function of the magnetic field for the FM Au–Al–Gd 1/1 ACs. The experimental data are fitted by the equation, $|\Delta S_M| = A\mu_0(H)^n$, where *A* and *n* are the fitting parameters. The values of the exponent *n* are to be 0.71, 0.63, 0.62, 0.67, 0.74, and 0.85 for *e*/*a* = 1.60, 1.64, 1.68, 1.76, 1.84, and 1.86, respectively.

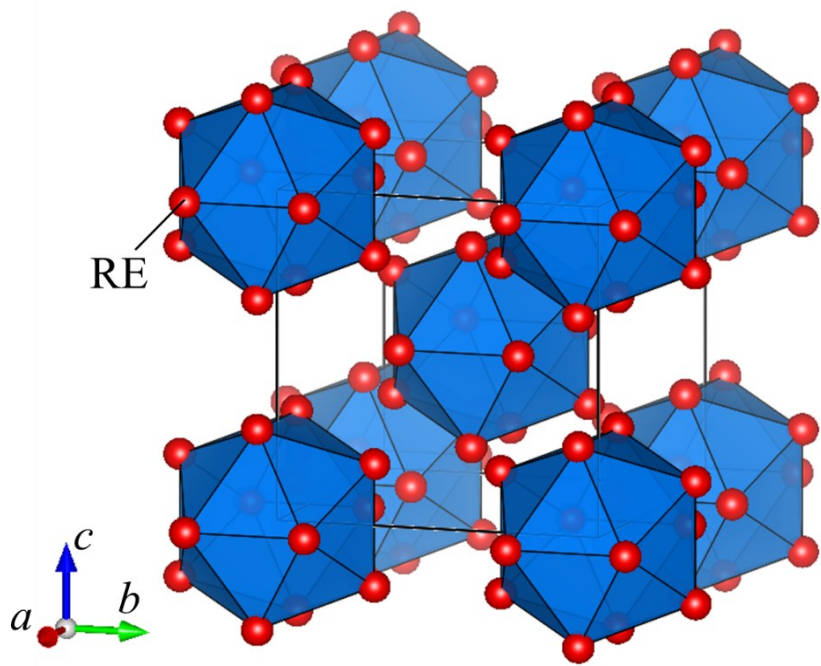

**Fig. S1.** The arrangement of rare-earth (RE) atoms in the Tsai-type 1/1 ACs, being visualized by using the VESTA 3 software [K. Momma and F. Izumi, J. Appl. Crystallogr. **44**, 1272 (2011).].

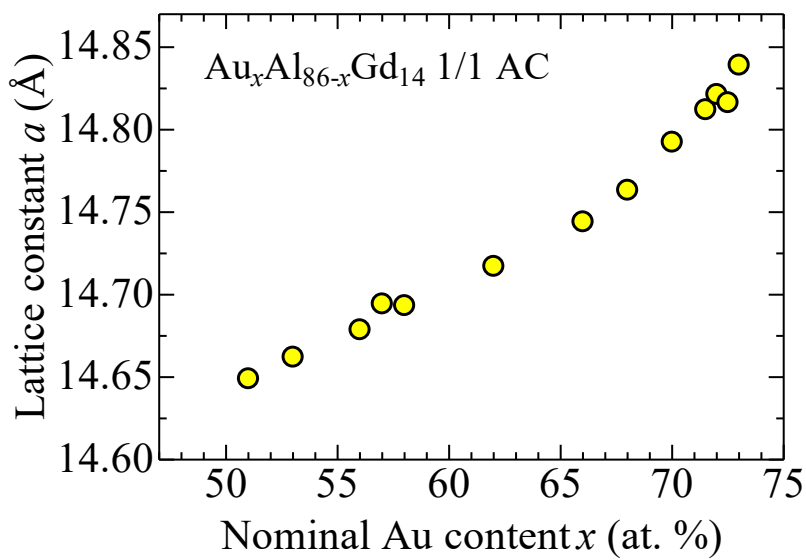

**Fig. S2.** Variation of lattice constant ($a$) with nominal Au content $x$ in the synthesized $Au_xAl_{86-x}Gd_{14}$ ($x$ = 51–73) samples.

**Table S1.** Nominal and analyzed compositions of the synthesized samples.

| Nominal compositions | Analyzed compositions |
|---|---|
| $Au_{51}Al_{35}Gd_{14}$ | $Au_{53.8}Al_{29.3}Gd_{16.9}$ |
| $Au_{53}Al_{33}Gd_{14}$ | $Au_{56.3}Al_{27.1}Gd_{16.6}$ |
| $Au_{56}Al_{30}Gd_{14}$ | $Au_{60.1}Al_{23.7}Gd_{16.2}$ |
| $Au_{57}Al_{29}Gd_{14}$ | $Au_{60.3}Al_{23.1}Gd_{16.6}$ |
| $Au_{58}Al_{28}Gd_{14}$ | $Au_{61.2}Al_{22.7}Gd_{16.1}$ |
| $Au_{62}Al_{24}Gd_{14}$ | $Au_{65.0}Al_{19.1}Gd_{16.0}$ |
| $Au_{66}Al_{20}Gd_{14}$ | $Au_{68.5}Al_{16.3}Gd_{15.6}$ |
| $Au_{68}Al_{18}Gd_{14}$ | $Au_{68.9}Al_{15.0}Gd_{16.1}$ |
| $Au_{70}Al_{16}Gd_{14}$ | $Au_{71.9}Al_{11.9}Gd_{16.2}$ |
| $Au_{71.5}Al_{14.5}Gd_{14}$ | $Au_{72.5}Al_{12.2}Gd_{15.3}$ |
| $Au_{72.0}Al_{14.0}Gd_{14}$ | $Au_{73.7}Al_{11.1}Gd_{15.2}$ |
| $Au_{72.5}Al_{13.5}Gd_{14}$ | $Au_{73.7}Al_{10.8}Gd_{15.5}$ |
| $Au_{73.0}Al_{13.0}Gd_{14}$ | $Au_{75.2}Al_{9.3}Gd_{15.5}$ |

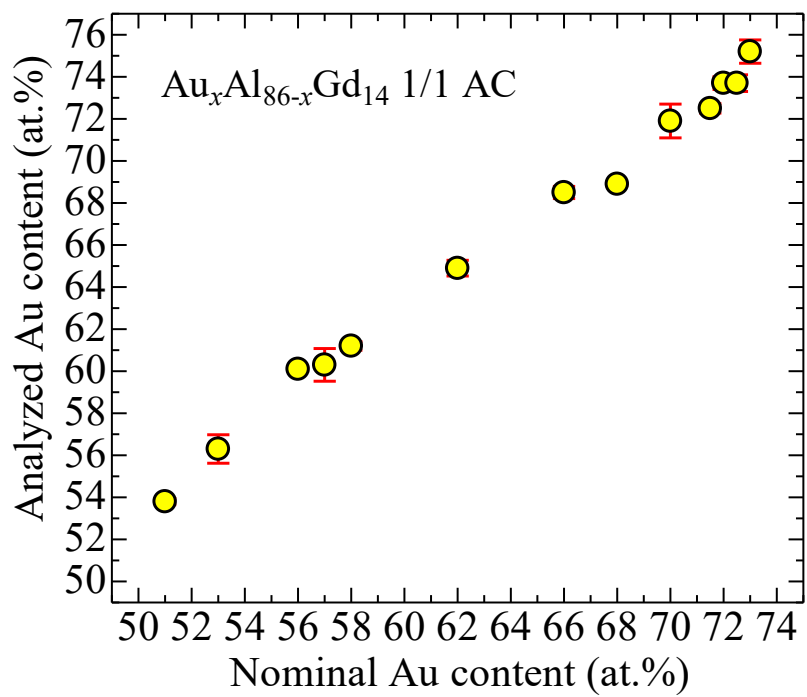

**Fig. S3.** The analyzed (by SEM-EDX analysis) Au content plotted as a function of the nominal Au content $x$ in the synthesized $Au_xAl_{86-x}Gd_{14}$ ($x$ = 51–73) samples.

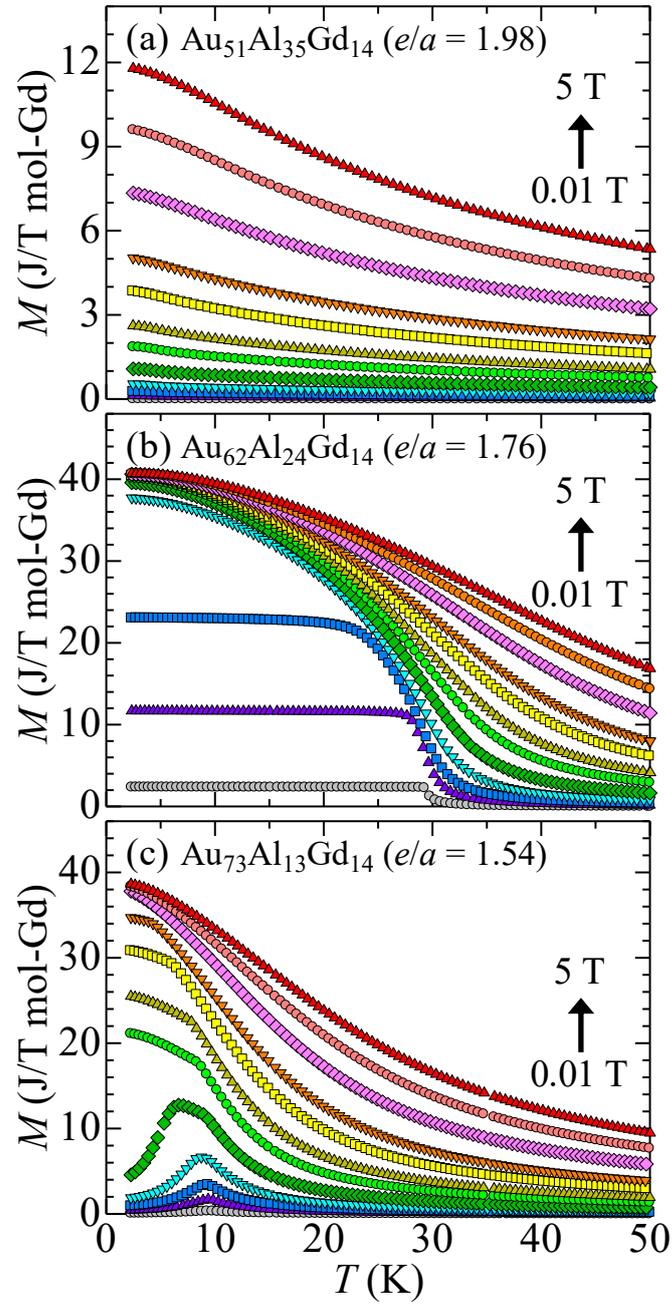

**Fig. S4.** (a)–(c) M-T curves of (a) $Au_{51}Al_{35}Gd_{14}$, (b) $Au_{62}Al_{24}Gd_{14}$, and (c) $Au_{73}Al_{13}Gd_{14}$ at various magnetic field (0.01, 0.05, 0.10, 0.20, 0.40, 0.70, 1.0, 1.5, 2.0, 3.0, 4.0, and 5.0 T) under FC processes.

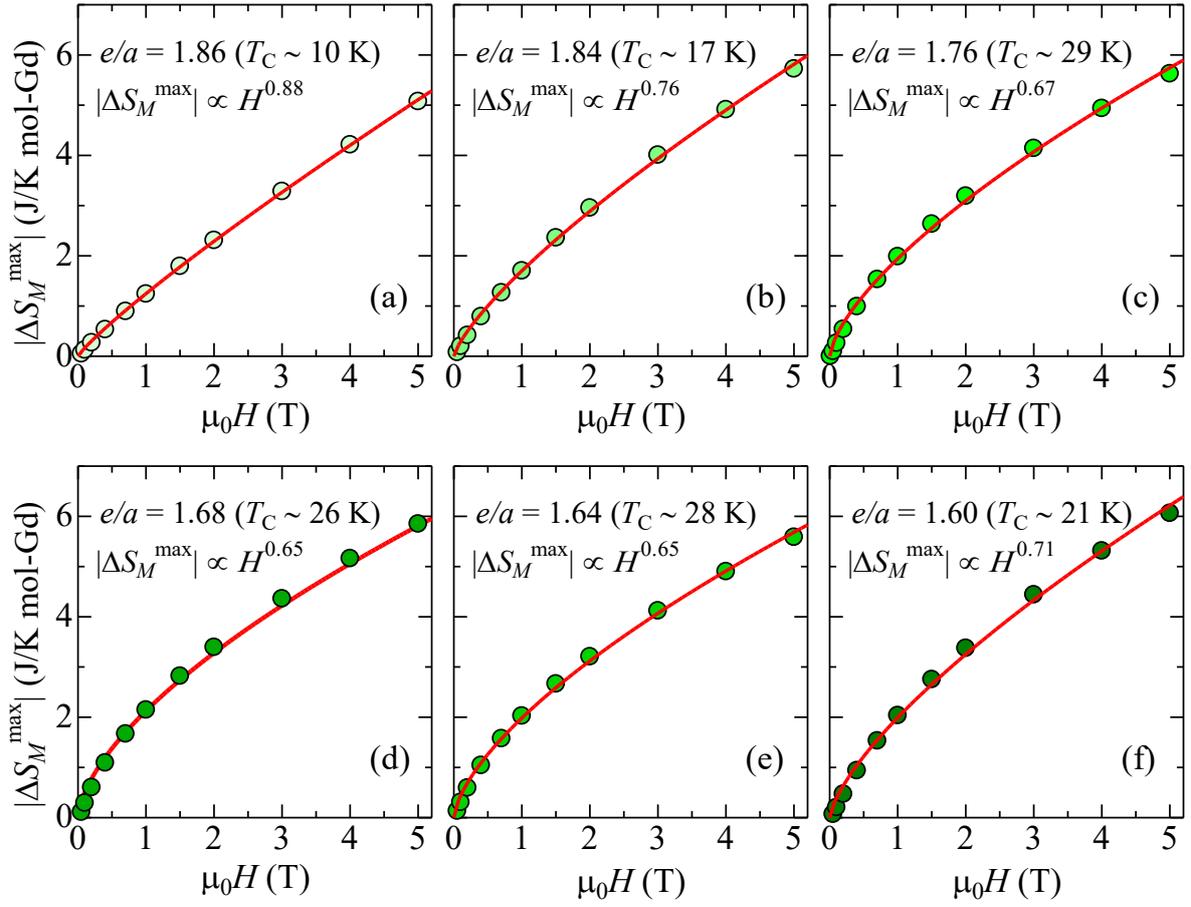

**Fig. S5.** (a)–(f) The $|\Delta S_M^{\max}|$ as a function of magnetic field in FM Au–Al–Gd 1/1 ACs. The lines indicate the fitting results.